\documentclass[11pt, oneside]{amsart}
\usepackage[utf8]{inputenc}
\usepackage[left=0.8in, right=0.8in, top=0.8in, bottom=0.9in, centering]{geometry}      
\usepackage{float, epsfig, natbib, lineno, appendix}
\usepackage{color, amssymb, amsmath, amsthm, verbatim, wasysym, mathrsfs}
\usepackage[mathscr]{eucal}
\usepackage{hyperref} 
\usepackage[font={footnotesize}]{caption}
 \usepackage{amsaddr}
\usepackage{scalerel, stackengine}
\setstackEOL{\#}
\stackMath
\def\hatgap{2pt}
\def\subdown{-2pt}
\newcommand\reallywidehat[2][]{ \renewcommand\stackalignment{l} \stackon[\hatgap]{#2}{ \stretchto{
    \scalerel*[\widthof{$#2$}]{\kern-.6pt\bigwedge\kern-.6pt}
    {\rule[-\textheight/2]{1ex}{\textheight}}}
    {0.5ex}_{\smash{ \belowbaseline[\subdown]{\scriptstyle#1} }}
}}

\newcommand{\beq}		{\begin{equation}}
\newcommand{\eeq}		{\end{equation}}

\begin{document}
\title{Two-dimensional flow on the sphere}

\author{Rick Salmon and Nick Pizzo}
\address{Scripps Institution of Oceanography, University of California, San Diego}
\email{Correspondence: rsalmon@ucsd.edu}

    

\maketitle


\begin{abstract}{Equilibrium statistical mechanics predicts that inviscid, two-dimensional, incompressible flow on the sphere eventually reaches a state in which spherical harmonic modes of degrees $n=1$ and $n=2$ hold all the energy. By a separate theory, such flow is static in a reference frame rotating at angular speed $2\Omega/3$ with respect to the inertial frame.  The vorticity field in the static frame is an accident of the initial conditions, but, once established, it lasts forever under the stated assumptions.  We investigate the possibility of such behavior with a stereographic-coordinate model that conserves energy and enstrophy when the viscosity vanishes.}
\end{abstract}



\section{Introduction}

Jack Herring made fundamental contributions to the theory of two-dimensional turbulence and to quasigeostrophic turbulence.  The former is the prototype of the latter, which adds the key dynamical ingredients---rotation, density stratification, and topography---required to make the theory relevant to the Earth's atmosphere and oceans.  Much of the theory is based based upon planar geometry and the hope that the beta-plane approximation captures the most significant effects of curvature and coordinate system rotation.  A primary purpose of this paper will be to show that this is not the case: Two-dimensional incompressible flow on the sphere differs significantly from two-dimensional flow on the plane even in the case of vanishing rotation, or, as we prefer to say, even in the case of vanishing angular momentum.

A second primary purpose is to illustrate the advantages of stereographic coordinates for both analytic and numerical work on the sphere.  In stereographic coordinates the equations governing two-dimensional turbulence on the sphere take a form, (\ref{27}) or (\ref{32}-\ref{34}), that is very similar to the corresponding formulation in Cartesian coordinates on the plane. Only the appearance of the smoothly varying metric coefficient $h$ distinguishes the two dynamics.  From a mathematical point of view, $h$ is responsible for all of the differences between flow in the two geometries.

\begin{figure}[H]
\includegraphics[width=5. cm]{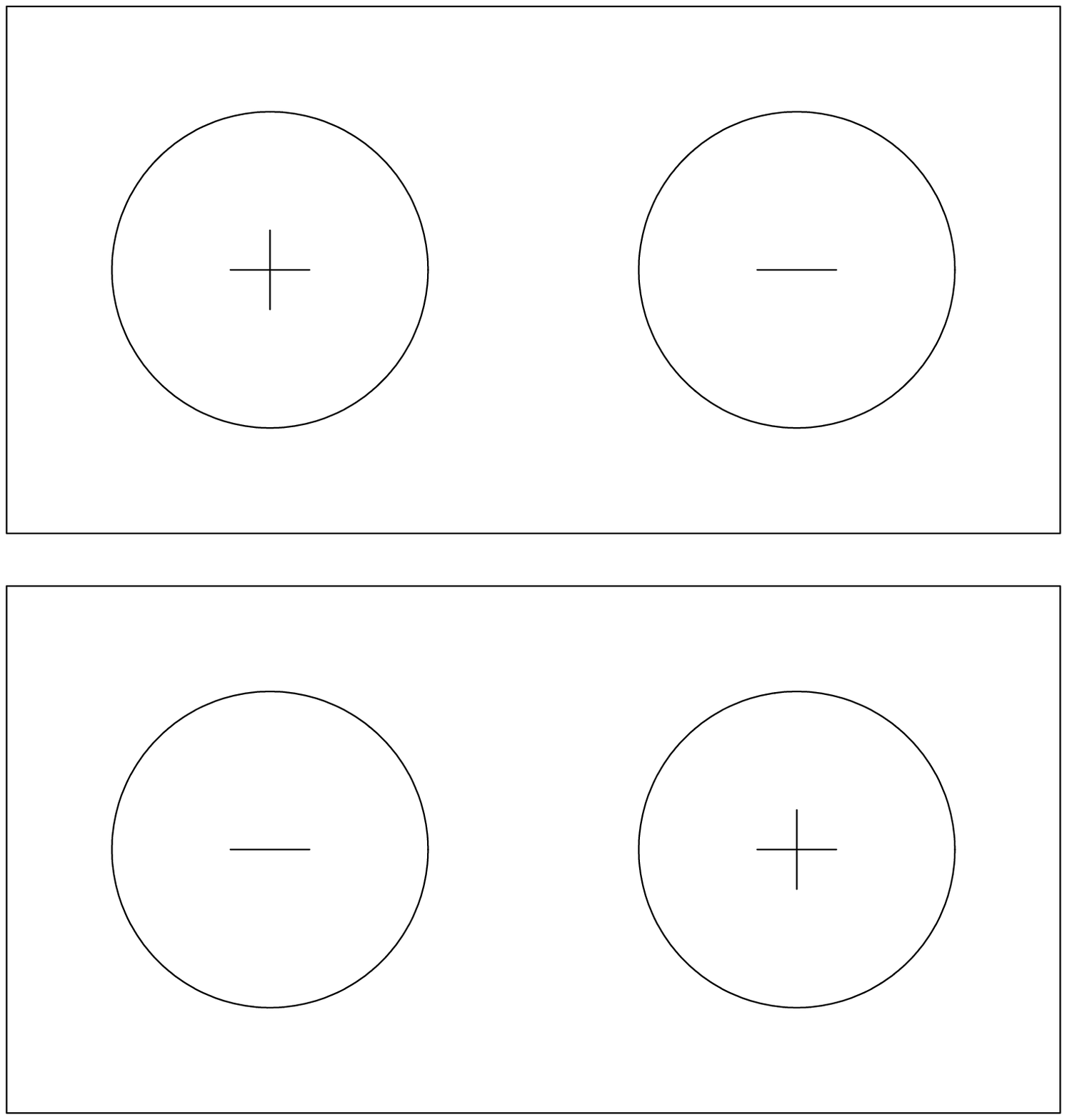}
\caption{If all the energy ends up in the lowest wavenumber mode of the rectangular domain, it must occupy one of the two states depicted.\label{fig1}}
\end{figure}

To quickly appreciate the effects of spherical geometry, first consider inviscid, incompressible, two-dimensional flow on the plane.  Unlike on the sphere, which has no boundary, the prescribed boundary on the plane is determinative. Let it be the rectangle shown in Figure \ref{fig1}.

Let the flow within this rectangle be truncated to exclude all the Fourier modes with wavenumbers greater than some prescribed cutoff $k_c$.  Let the initial energy be concentrated at a wavenumber $k_0$ that lies between $k_c$ and the lowest wavenumber as determined by the boundary of the system.

Equilibrium statistical mechanics \citep{Kraichnan1967} predicts the statistically steady state attained by this system after an undetermined period of adjustment.  Interest attaches to the sequence of equilibrium states that occur as $k_c \to \infty$.  In this limit the theory predicts that \emph{all} of the energy ends up in the lowest-wavenumber mode, and that all of the enstrophy not contained in that mode appears in modes near $k_c$.  As $k_c \to \infty$ this excess enstrophy is expelled to infinity, disappearing in much the same way as if it had been removed by viscosity.  The tendency for all the energy to crowd into a single lowest mode is often referred to as `Bose-Einstein condensation', after a closely analogous result in quantum mechanics.

Suppose that the initial conditions are such that the conserved circulation around the rectangular boundary vanishes.  Then the lowest-wavenumber mode, which has a single extremum within the rectangle, remains unexcited.  All of the energy flows into the next-lowest mode, the one depicted in Figure \ref{fig1}.  Because the boundary is a rectangle (and not, say, a square or a circle), there is only one next-lowest mode.  If all of the energy ends up in this mode, it must appear in one of the two states depicted in Figure \ref{fig1}. Moreover, once either of these states is established, the system cannot flip suddenly to the opposite state, because that would require the energy in one of these states to flow temporarily into higher-wavenumber modes, and that, according to statistical mechanics, is extremely unlikely.  Thus the two states depicted in Figure \ref{fig1} act as a single-register binary memory.  The state that actually occurs is an accident of the initial conditions, but once established it lasts for a very long time.

Still considering flow on the plane, suppose that there are \emph{two} lowest modes available to receive all the energy.  For the statistical mechanics we take the microcanonical ensemble as seems appropriate.  Let the energy be $E=x^2+y^2$ where $x$ and $y$ are the amplitudes of the two modes. If $x$ were the only mode, then, as already noted, its probability distribution would be
\begin{equation}
P(x)=\delta(x-\sqrt{E}) + \delta(x+\sqrt{E})
\label{01}
\end{equation}
where, here and below, we ignore normalization constants.  However, if there are two modes, then the probability distribution of either mode is
\begin{equation}
P(x)=\int_{-\infty}^{+\infty} dy \; \delta(x^2+y^2-E) \propto \frac{1}{\sqrt{E-x^2}}
\label{02}
\end{equation}
if $x^2<E$ and zero otherwise.  Like (\ref{01}), the distribution (\ref{02}) is sharply peaked at $\pm \sqrt{E}$ but now there is significant probability of finding $x$ anywhere in between.  If there are three lowest modes, then the analogous calculation predicts that $x$ is uniformly distributed between  $\pm \sqrt{E}$.  The implication of these calculations is clear:  If the lowest available mode is degenerate with two or more members, then each member sees the other members as a reservoir from which energy may be borrowed or loaned, allowing leakage between states and a breakdown of memory.  In all of these examples the average flow predicted by statistical mechanics is misleading and beside the point.  For example, if the ensemble consists of the two equal-probability states depicted in Figure \ref{fig1}, then the average flow vanishes, but a vanishing velocity field is very unrepresentative of either state.  Such considerations demonstrate how even the simplest calculations with probability distributions require careful interpretation.

Now consider the corresponding equilibrium state on the sphere \citep{Frederiksen80}.  Spherical harmonics replace Fourier modes, and the harmonic degree $n$ with cutoff $n_c$ replaces the wavenumber $k$ with cutoff $k_c$.  The lowest mode $Y_0^0$ is irrelevant because the stream function is arbitrary by a constant, and because the vorticity integrates to zero over the sphere (Gauss's constraint).  The next lowest modes are three: $Y_1^0$, $Y_1^{\pm 1}$.  However, the amplitudes of these three modes are determined by the three components of angular momentum, which is conserved by our dynamics even with nonzero viscosity.  Thus the lowest-degree modes into which the energy can pile up are those of degree $n=2$, namely, $Y_2^0$, $Y_2^{\pm 1}$, $Y_2^{\pm 2}$.  By the reasoning of the previous paragraph, the Bose-Einstein condensate would be one in which these 5 modes freely exchange energy.  Instead, it turns out, the 5 modes of degree 2 lock together in a pattern that is perfectly static in a uniformly rotating reference frame.  This behavior is dramatically different from flow on the plane, and is solely a consequence of spherical geometry.

Let the sphere be centered at $x=y=z=0$, and suppose, without loss of generality, that the angular momentum (if nonzero) points toward positive $z$.  Then the amplitude of the $Y_1^0$-mode is a nonzero constant, and the amplitudes of $Y_1^{\pm 1}$ vanish.  The $Y_1^0$-mode corresponds to a solid-body flow about the $z$-axis with constant angular speed $\Omega$, and the corresponding vorticity is called the Coriolis parameter.  As shown in Section 3, the flow consisting of the $Y_1^0$-mode and the 5 $Y_2$-modes with arbitrary amplitudes represents an exact solution to the inviscid vorticity equation on the sphere.  If $\Omega=0$ this flow field is steady in the inertial frame.  If $\Omega\neq 0$ the flow field is steady in a reference frame rotating at speed $2\Omega/3$ with respect to the inertial frame.
The amplitude $\Omega$ of the $Y_1^0$-mode is determined by the conserved angular momentum.  The amplitudes of the $Y_2$-modes are accidents of the initial conditions in the same way that initial conditions determine which of the two states in Figure \ref{fig1} eventually emerge.  However, once these amplitudes are determined via Bose-Einstein condensation, they persist indefinitely under the stated assumptions. Thus Bose-Einstein condensation into $Y_2$-modes represent a potential memory storage that does not, as in planar geometry, require a `probability well' for its preservation. 
For the plane, as for the sphere, triads with the same wavenumber magnitude (the analog of spherical harmonic degree) do not interact, either with each other or with modes of different wavenumber magnitude.  However, typical planar geometries forbid the occurrence of many modes with the same wavenumber magnitude.  For instance, in the case of an infinitely periodic box, in which the wave vector corresponds to integer pairs $(n,m)$, modes with the same $n^2+m^2$ are rare, as is seen from a list of Pythagorean Triples.

More generally, flow consisting only of $Y_1$ modes and modes of a single spherical harmonic degree $n$ is perfectly static in a reference frame rotating at angular speed
\begin{equation}
\Omega-\frac{2\Omega}{n(n+1)}
\label{03}
\end{equation}
with respect to the inertial frame. According to \cite{Kochin64} this result was known to Ertel.  Section 3 supplies details.
There are $2n+1$ spherical harmonics with the same degree $n$. These facts invite us to think of the stream function for flow on the sphere as the sum,
\begin{equation}
\psi(\xi,\eta,t)=\sum_{n=1}^{\infty} \psi_n(\xi,\eta,t)
\end{equation}
of contributions $\psi_n$ of degree $n$, where each $\psi_n$ is itself the sum of $2n+1$ spherical harmonics.  The coordinates $(\xi,\eta)$ are arbitrary, but we strongly prefer them to be stereographic coordinates.
Again, if only $\psi_1$ and one $\psi_n$ with $n>1$ are present, then $\psi$ is steady in the frame rotating at angular speed (\ref{03}).  In the general case where all $\psi_n$ are present, the various $\psi_n$ interact to produce a turbulent flow.  However, the interaction between $\psi_n$ with very different autorotation rates (\ref{03})---very different $n$---is likely to be weak.  And since the difference between autorotation rates is greatest and varies most rapidly at small $n$, we expect energy transfer to large spatial scales to slow dramatically as the energy reaches low $n$, for sufficiently large $\Omega$.  This is essentially \cite{Rhines75}'s  `beta-arrest' phenomenon for flow on the beta plane, but now taking a form particular to the sphere.  See also \cite{Vallis93}.

This study was begun with the hope that the special static solutions described above might be a factor in intermediate-range weather prediction.  Although flickers of that hope remain, there are significant differences between the Earth's atmosphere and the very idealized dynamics considered in this paper.  
First and foremost, angular momentum conservation applies only to the entire Earth-atmosphere-ocean system. Mountain torque provides the main coupling between the atmosphere and the solid planet.  The latter, with its enormously greater moment of inertia, experiences only small changes in its rotation rate, but fluctuations in the  $Y_1^0$ component of atmospheric flow are very significant and lead to a peak in the atmospheric energy spectrum at $n=1$ \citep{Boer83}.
Second, because the Rhines scale corresponding to the Earth's atmosphere is smaller than the planetary radius, beta-arrest prevents the leftward movement of energy through the spectrum from reaching $n=2$.  As a result of these two factors, the atmospheric energy spectrum actually shows a minimum at $n=2$. See \cite{Sawford83} for the statistical mechanics including mountain torque and \cite{Egger2007} for a thorough review of atmospheric angular momentum.

The plan of the paper is as follows.  In Section 2 and Appendix A we derive the Navier Stokes equations for flow on the sphere in stereographic coordinates.  
The form of Navier-Stokes viscosity for flow on curved surfaces has been the subject of disagreement. For instance, the viscosity given by \cite{Gill82} does not conserve angular momentum.  
We follow the recommendation of \cite{Gilbert2014} that the viscosity conserve angular momentum, and in Appendix A we show that their recommended viscosity takes an especially simple form in stereographic coordinates.  

In Section 3 we prove the special solutions described briefly above.  Our proof using stereographic coordinates is substantially simpler than the proofs given by \cite{Thompson82} or \cite{Verkley84} using spherical coordinates.

In Section 4 and Appendix B we use the method of \cite{Salmon89} to construct an Arakawa-type model in stereographic coordinates.  The model conserves energy and enstrophy when the viscosity vanishes.  

Section 5 demonstrates the accuracy of the model and tests our speculation that Bose-Einstein condensation on the sphere produces a static flow.  Our results suggest that the time required to achieve the static state is probably infinite, but that the flow rather quickly becomes quasi-static.

It is a pleasure to dedicate this paper to the memory of Jack Herring.  Rick Salmon acknowledges the importance of Jack's mentorship at an early stage of his career and holds the happy memories of a lifelong friendship.  Rick cannot think of Jack without also remembering his wife Betty, who shared Jack's kind and generous spirit, and greatly enjoyed the foibles and eccentricities of his scientific friends.

\section{Sterographic coordinates}

We solve the vorticity equation for incompressible flow on the unit sphere,
\begin{equation}
x^2+y^2+z^2=1.
\label{1}
\end{equation}  
We refer to the point $(x,y,z)=(0,0,1)$ as the `north pole', and $(x,y,z)=(0,0,-1)$ as the `south pole', even when the angular momentum vanishes, the case commonly referred to as `non-rotating.'  The `equator' is the intersection of the plane $z=0$ with (\ref{1}). The conserved angular momentum is a vector in the direction of $(0,0,1)$.

Besides the Cartesian embedding coordinates $(x,y,z)$, we shall refer to three other coordinate systems.  The first of these are the usual spherical coordinates, defined by
\begin{align}
x&=\cos \theta \cos \lambda
\notag \\
y&=\cos \theta \sin \lambda
\notag \\
z&=\sin \theta
\label{2}
\end{align}
where $\lambda$ is the longitude and $\theta$ the latitude.
The $(\lambda,\theta)$ coordinates cover the sphere except for singularities at the two poles.
We also use the stereographic coordinates
\begin{equation}
\xi=\frac{x}{1-z}, \;\;\;\;  \eta=\frac{y}{1-z}
\label{3}
\end{equation}
defined by a line, emanating from the north pole, that intersects the equatorial plane at $(\xi,\eta,0)$ and the sphere at $(x,y,z)$.  $(\xi,\eta)$ cover the sphere except for a singularity at infinity, corresponding to the north pole.
We also use
\begin{equation}
\hat{\xi}=\frac{x}{1+z}, \;\;\;\;  \hat{\eta}=\frac{y}{1+z}
\label{4}
\end{equation}
defined by the intersection with the equatorial plane of a line emanating from the south pole.   $(\hat{\xi},\hat{\eta})$ cover the sphere except for singularity at infinity corresponding to the south pole.  The stereographic systems (\ref{3}) and (\ref{4}) prove to be more useful than the spherical coordinates; the primary challenge is to match them together.   Our strategy will be to solve the southern hemisphere dynamics in $(\xi,\eta)$ within the unit circle
\begin{equation}
\xi^2+\eta^2<1,
\label{5}
\end{equation}
 to solve the northern hemisphere dynamics in $(\hat{\xi},\hat{\eta})$  within the unit circle
\begin{equation}
\hat{\xi}^2+\hat{\eta}^2<1,
\label{6}
\end{equation}
and to match the two solutions together at the equator,
\begin{equation}
\xi^2+\eta^2=\hat{\xi}^2+\hat{\eta}^2=1.
\label{7}
\end{equation}
The transformation equations between the two stereographic systems are
\begin{equation}
(\xi,\eta)=\frac{1}{\hat{r}^2} (\hat{\xi},\hat{\eta}), \;\;\;\;
(\hat{\xi},\hat{\eta})=\frac{1}{r^2} (\xi,\eta)
\label{8}
\end{equation}
where
\begin{equation}
r^2\equiv \xi^2+\eta^2, \;\;\;\; \hat{r}^2\equiv (\hat{\xi})^2+(\hat{\eta})^2.
\label{9}
\end{equation}
A useful relation is
\begin{equation}
r \hat{r}=1.
\label{10}
\end{equation}
Points within the unit circle on the $(\hat{\xi},\hat{\eta})$ plane transform to points outside the unit circle on the $(\xi,\eta)$ plane, and vice versa.  On the equator itself, $(\xi,\eta)=(\hat{\xi},\hat{\eta})$. It is also useful to record the inverse transformations from either set of stereographic coordinates to the Cartesian embedding coordinates,
\begin{align}
x&=\frac{2\xi}{\xi^2+\eta^2+1}
=\frac{2\hat{\xi}}{\hat{\xi}^2+\hat{\eta}^2+1}
\label{11}\\
y&=\frac{2\eta}{\xi^2+\eta^2+1}
=\frac{2\hat{\eta}}{\hat{\xi}^2+\hat{\eta}^2+1}
\label{12}\\
z&=\frac{\xi^2+\eta^2-1}{\xi^2+\eta^2+1}
=\frac{1-\hat{\xi}^2-\hat{\eta}^2}{\hat{\xi}^2+\hat{\eta}^2+1}.
\label{13}
\end{align}
For a thorough introduction to stereographhic coordinates, see \cite{Needham1997,Needham2021}.

Infinitesimal displacements on the surface of the sphere satisfy
\begin{align}
ds^2=dx^2+dy^2+dz^2&=\cos^2\theta d\lambda^2+d\theta^2
\notag \\
&=\frac{4}{\left( 1+ \xi^2+\eta^2 \right)^2}\left( d\xi^2 + d\eta^2 \right)
\notag \\
&=\frac{4}{\left( 1+ \hat{\xi}^2+\hat{\eta}^2 \right)^2}\left( d\hat{\xi}^2 + d\hat{\eta}^2 \right)
\label{14}
\end{align}
where $ds$ is infinitesimal distance tangent to the surface of the sphere.  Thus all of our systems fit the general form
\begin{equation}
ds^2=h_1(\xi_1,\xi_2)^2 (d\xi_1)^2+ h_2(\xi_1,\xi_2)^2 (d\xi_2)^2
\label{15}
\end{equation}
where $(\xi_1,\xi_2)$ are general orthogonal coordinates, and $\mathrm{diag}(h_1^2,h_2^2)$ is the metric tensor.
In arbitrary orthogonal coordinates, the incompressibility condition takes the form
\begin{equation}
\frac{\partial}{\partial \xi_1} \left( h_1 h_2 \dot{\xi}_1 \right)
+
\frac{\partial}{\partial \xi_2} \left( h_1 h_2 \dot{\xi}_2 \right) = 0
\label{16}
\end{equation}
where the overdot denotes the time derivative following a fluid particle.  Since the velocity components tangent to the sphere in the directions of $\xi_1$ and $\xi_2$ are respectively
\begin{equation}
U_1=h_1 \dot{\xi}_1 , \;\;\;\; U_2=h_2 \dot{\xi}_2
\label{17}
\end{equation}
(\ref{16}) is equivalent to
\begin{equation}
\frac{\partial}{\partial \xi_1} \left( h_2 U_1 \right)
+
\frac{\partial}{\partial \xi_2} \left( h_1 U_2 \right) = 0
\label{18}
\end{equation}
which implies
\begin{equation}
U_1=-\frac{1}{h_2} \frac{\partial \psi}{\partial \xi^2}, \;\;\;\;
U_2=+\frac{1}{h_1} \frac{\partial \psi}{\partial \xi^1}
\label{19}
\end{equation}
where $\psi(\xi_1,\xi_2,t)$ is the stream function.  The vorticity equation takes the form
\begin{equation}
q_t+\frac{1}{h_1 h_2} \frac{\partial(\psi,q)}{\partial(\xi_1,\xi_2)}=0
\label{20}
\end{equation}
where
\begin{equation}
q= \nabla_{LB}^2 \psi
\label{21}
\end{equation}
is the vorticity, and
\begin{equation}
\nabla_{LB}^2 \psi  \equiv
\frac{1}{h_1 h_2} \left(
\frac{\partial}{\partial \xi_1}   
\left( \frac{h_2}{h_1}\frac{\partial \psi}{\partial \xi_1} \right)
+
\frac{\partial}{\partial \xi_2}
\left( \frac{h_1}{h_2}\frac{\partial \psi}{\partial \xi_2} \right)
\right)
\label{22}
\end{equation}
is the Laplace-Beltrami operator.

In the case of spherical coordinates $(\xi_1,\xi_2)=(\lambda,\theta)$, $(h_1,h_2)=(\cos \theta, 1)$,
and (\ref{19}-\ref{21}) take the familiar forms
\begin{equation}
U_\lambda=- \frac{\partial \psi}{\partial \theta}, \;\;\;\;
U_\theta=+\frac{1}{\cos \theta} \frac{\partial \psi}{\partial \lambda}
\label{23}
\end{equation}
\begin{equation}
q_t+\frac{1}{\cos \theta} \frac{\partial(\psi,q)}{\partial(\lambda,\theta)}=0
\label{24}
\end{equation}
and
\begin{equation}
q=\frac{1}{\cos^2 \theta} \frac{\partial^2 \psi}{\partial \lambda^2}
+\frac{1}{\cos \theta} \frac{\partial }{\partial \theta}
\left( \cos \theta  \frac{\partial \psi}{\partial \theta} \right)
\equiv \nabla_{LB}^2 \psi.
\label{25}
\end{equation}
  
To obtain the dynamics in stereographic coordinates, we proceed from the general equations (\ref{20}-\ref{21}). Some care must be taken concerning the handedness of the coordinate systems.  Viewed from the exterior of the unit sphere, $(\hat{\xi},\hat{\eta})$ and $(\lambda,\theta)$ constitute right-handed systems, whereas $(\xi,\eta)$ constitutes a left-handed system.  That is
\begin{equation}
\frac{\partial(\hat{\xi},\hat{\eta})}{\partial(\lambda,\theta)}>0,  \;\;\;\;
\frac{\partial(\xi,\eta)}{\partial(\lambda,\theta)}<0.
\label{26}
\end{equation}
The general covariant point of view cares nothing about handedness, but we want our computed vorticity and stream function to be the same as would be obtained by solving the entire problem in spherical coordinates, and, in particular, we want these quantities to be continuous at the equator;  it would be a great nuisance to keep track of differently defined stream functions in the two hemispheres.  With this in mind, the northern hemisphere equations take the forms
\begin{equation}
\hat{h}^2 q_t+ \frac{\partial(\psi,q)}{\partial(\hat{\xi},\hat{\eta})}=0, \;\;\;\;
\hat{h}^2 q
= \hat{\nabla}^2 \psi, \;\;\;\;
\hat{h}=\frac{2}{1+\hat{\xi}^2+\hat{\eta}^2}
\label{27}
\end{equation}
within the unit circle (\ref{6}); and the southern hemisphere equations take the forms
\begin{equation}
h^2 q_t- \frac{\partial(\psi,q)}{\partial(\xi,\eta)}=0, \;\;\;\;
h^2q
= \nabla^2 \psi, \;\;\;\;
h=\frac{2}{1+\xi^2+\eta^2}
\label{28}
\end{equation}
within the unit circle (\ref{5}).  Here,
\begin{equation}
\nabla^2 \psi \equiv
\frac{\partial^2 \psi }{\partial \xi^2} + \frac{\partial^2 \psi}{\partial \eta^2}
, \;\;\;\;
\hat{\nabla}^2 \psi \equiv
\frac{\partial^2 \psi }{\partial \hat{\xi}^2} + \frac{\partial^2 \psi}{\partial \hat{\eta}^2}.
\label{29}
\end{equation}
The Laplace Beltrami operator in (\ref{22}) or (\ref{25}) is equivalent to $\hat{h}^{-2}\hat{\nabla}^2$ and $h^{-2}\nabla^2$.
Apart from the factors of $h$ and $\hat{h}$, (\ref{27}-\ref{29}) have the same form as the corresponding equations for planar geometry in Cartesian coordinates.  This greatly facilitates numerical solution.
The sign change between (\ref{27}) and (\ref{28}) reflects the change in handedness.  To see that this makes sense, recall that the first equation in (\ref{27}) and (\ref{28}) also applies to a passive scalar advected by the velocity field represented by $\psi$.  The sign change between (\ref{27}) and (\ref{28}) ensures that the (cross-equatorial) flux of $q$ out of the unit circle (\ref{5}) matches the flux of $q$ into the unit circle (\ref{6}), with no need to redefine or re-interpret $\psi$ or $q$.

Alternatively, by replacing the vorticity $q$ in (\ref{27}) and (\ref{28}) by 
\begin{equation}
q+f \equiv q+2\Omega\sin \theta=q+2\Omega z
=q+2\Omega \frac{\xi^2+\eta^2-1}{\xi^2+\eta^2+1}
=q+2\Omega \frac{1-\hat{\xi}^2-\hat{\eta}^2}{\hat{\xi}^2+\hat{\eta}^2+1}
\label{29a}
\end{equation}
we obtain the equations of motion in a frame rotating at angular speed $\Omega$.

We solve (\ref{27}) on (\ref{6}), and (\ref{28}) on (\ref{5}), with the matching conditions that $\psi$ and $q$ be continuous at the equator (\ref{7}).  If the initial condition is given by prescribing $q$, then $q$ must satisfy the Gauss constraint, 
\begin{equation}
\iint d\xi d\eta \; h^2 q + \iint d\hat{\xi} d\hat{\eta} \; \hat{h}^2 q 
= 0
\label{30}
\end{equation}
where the integrals are over the two unit circles (i.e. the whole sphere),  
but the constraint (\ref{30}) is automatically maintained thereafter.  If the initial condition is given by prescribing $\psi$, the Gauss constraint is automatically satisfied.

Although we work primarily with (\ref{27}) and (\ref{28}), it is useful to record the corresponding forms of the momentum equations.  Introducing the notation,
\begin{equation}
(u,v) \equiv h^2 (\dot{\xi},\dot{\eta})
\label{31}
\end{equation}
the southern hemisphere equations take the forms
\begin{align}
\frac{\partial u}{\partial t} -q v &= - \frac{\partial p}{\partial \xi} 
\label{32} \\
\frac{\partial v}{\partial t} +q u &= - \frac{\partial p}{\partial \eta}
\label{33} \\
\frac{\partial u}{ \partial \xi} + \frac{\partial v}{ \partial \eta}&=0
\label{34}\\
q&=\frac{1}{h^2} \left( 
\frac{\partial v}{\partial \xi} -
\frac{\partial u}{\partial \eta} \right).
\label{35}
\end{align}
The northern hemisphere equations take the same form as (\ref{32}-\ref{35}) but with hats applied to all the variables except $q$.  Again, if $h^2=1$, (\ref{32}-\ref{35}) are formally identical to the equations for two-dimensional incompressible motion in planar geometry.  However, no choice of variables can make $h^2$ equal unity on the unit sphere; instead it must obey the requirement
\begin{equation}
\frac{1}{h^4} \left[
\frac{\partial h}{\partial \xi_1} \frac{\partial h}{\partial \xi_1} +
\frac{\partial h}{\partial \xi_2} \frac{\partial h}{\partial \xi_2}
-h \left( \frac{\partial^2 h}{\partial \xi_1^2} + \frac{\partial^2 h}{\partial \xi_2^2} \right)
\right]=1
\label{36}
\end{equation}
that the Gaussian curvature be uniform and equal to unity.  The effects of curvature can never be transformed away, but it is interesting that they can be confined to a single factor in (\ref{35}).  This is what motivates the definition (\ref{31}).  However, it must be emphasized that the symbols $(u,v)$ invite misinterpretation: the velocity of fluid particles tangent to the sphere is 
\begin{equation}
(U,V)=h(\dot{\xi},\dot{\eta})=\frac{1}{h}(u,v)
\label{37}
\end{equation}
and \emph{not} $(u,v)$.

The inviscid dynamics (\ref{27}-\ref{28}) or (\ref{32}-\ref{35}) conserve energy in the form,
\begin{equation}
E=\frac{1}{2} \iint d\xi d\eta \; \nabla \psi \cdot \nabla \psi
\label{38}
\end{equation}
total vorticity in the form
\begin{equation}
\iint d\xi d\eta \; h^2 q = \iint d\xi d\eta \; \nabla^2 \psi \equiv 0
\label{39}
\end{equation}
and enstrophy in the form
\begin{equation}
Z=\frac{1}{2} \iint d\xi d\eta \; h^2 q^2 = \frac{1}{2} \iint d\xi d\eta \; h^{-2}(\nabla^2 \psi)^2.
\label{40}
\end{equation}
The integrations in (\ref{38}-\ref{40}) are over the whole sphere, with the understanding that they are actually computed as the sum of integrations over the two unit circles.  
The vanishing of (\ref{39}) corresponds to the Gauss constraint.  More generally, the inviscid  dynamics conserves Casimirs of the form
\begin{equation}
\iint d\xi d\eta \; h^2 F(q)
\label{40a}
\end{equation}
where $F(q)$ is any function of the vorticity for which the integral in (\ref{40a}) converges; (\ref{39}) and (\ref{40}) are particularly important cases of (\ref{40a}).

The inviscid dynamics also conserves angular momentum.  Angular momentum is a vector with 3 components, each of which is conserved.  However, since the location of the axis that determines the origin of our stereographic system has no particular significance, we may assume, without loss of generality, that the angular momentum vector points in the $z$-direction.  Then the sole nonvanishing component of angular momentum is
\begin{equation}
M=\iint d\xi d\eta \; h^2(\xi v - \eta u)= \iint d\xi d\eta \; h^3 q=\iint d\xi d\eta \; h \nabla^2 \psi.
\label{41}
\end{equation}

It remains to discuss the viscosity.  \cite{Gilbert2014} examine three proposed definitions of viscosity for incompressible flow on a curved surface.  All three are covariant, and none of the three can be rejected on purely fundamental grounds.  However, only one of the three forms of viscosity conserves angular momentum.  We agree with \cite{Gilbert2014} that this angular-momentum conserving property is the proper choice for applications in which the viscosity is actually an eddy viscosity representing the mixing---but not the destruction---of angular momentum by unresolved motions.  In Appendix A we show that the viscosity suggested by \cite{Gilbert2014} takes the surprisingly simple form
\begin{align}
\frac{\partial u}{\partial t} + \cdots &=  -\nu \frac{\partial q}{\partial \eta}
\label{42} \\
\frac{\partial v}{\partial t} + \cdots &=  +\nu \frac{\partial q}{\partial \xi}
\label{43}
\end{align}
in the dynamics (\ref{32}-\ref{35}), and
\begin{equation}
\hat{h}^2q_t + \cdots = \nu \hat{\nabla}^2 q, \;\;\;\; h^2q_t + \cdots = \nu \nabla^2 q
\label{44}
\end{equation}
in the dynamics (\ref{27}-\ref{28}).  The ellipses represent the conservative terms already discussed.  When the viscosity in (\ref{44}) is added to the dynamics (\ref{27}-\ref{28}) we find that
\begin{equation}
\frac{dE}{dt}=-\nu \iint d\xi d\eta \; h^2 q^2
\label{45}
\end{equation}
\begin{equation}
\frac{dZ}{dt}=-\nu \iint d\xi d\eta \; \nabla q \cdot \nabla q
\label{46}
\end{equation}
and
\begin{equation}
\frac{dM}{dt}=0.
\label{47}
\end{equation}
Thus viscosity dissipates energy and enstrophy, but not angular momentum.  The importance of angular momentum conservation by viscosity seems not to be widely appreciated.  

Many modelers prefer a more scale-selective hyperviscosity to any form of Navier-Stokes viscosity. However, hyperviscosity introduces unphysical effects. For example, hyperviscosity creates artificial local extrema in vorticity.  For this reason, we believe that Navier-Stokes viscosity is well worth the extra cost in spatial resolution.

We emphasize that, in our dynamics, the sphere plays no role except to shape the flow.  Without drag or mountain torque, only curvature and angular momentum conservation distinguish our dynamics from the flow in a box on the plane.

\section{Exact solutions}

Let $Y_n$ be any solution of
\begin{equation}
\nabla_{LB}^2 Y_n =-n(n+1) Y_n
\label{401}
\end{equation}
where $n$ is a positive integer.  The coordinate system used to represent $Y_n$ is arbitrary.  In a spherical coordinate system, the general solution of (\ref{401}) is
\begin{equation}
Y_n(\theta,\lambda)=\sum_{m=0}^n A_m P_n^m(\sin \theta) \cos \left( m\lambda + \gamma_m \right)
\label{402}
\end{equation}
where $P_n^m$ is a Legendre function of the first kind, and $A_m$ and $\gamma_m$ are arbitrary constants.  We recognize (\ref{402}) as the sum of all spherical harmonics of degree $n$ with arbitrary amplitudes $A_m$ and phases $\gamma_m$.  The customary normalization coefficients have been lumped into the $A_m$.  The representation (\ref{402}) applies to any system of spherical coordinates.  That is, if we transform (\ref{402}) to an alternative set of spherical coordinates--- obtained, for example, by changing the polar axis with respect to which the spherical harmonics are defined---we obtain an expression identical to (\ref{402}) but with a different set of $A_m$'s and $\gamma_m$'s.  We may in fact use any system of coordinates, including the more convenient stereographic coordinates, because the following proof relies only on the covariant property (\ref{401}).

Let $\psi_{(l,m,n)}^\Omega$ be the streamfunction corresponding to solid-body flow at angular speed $\Omega$ about an axis in the direction of the unit vector $(l,m,n)$.   
Let $Y_n$ be a solution of (\ref{401}), and let
$Y_n^{\omega}$ be the pattern $Y_n$ rotating at the angular speed
\begin{equation}
\omega=\Omega-\frac{2\Omega}{n(n+1)}
\label{403}
\end{equation}
about the axis $(l,m,n)$.  Then
\begin{equation}
\psi = \psi_{(l,m,n)}^\Omega + Y_n^{\omega}
\label{404}
\end{equation}
is a solution of our dynamics.

For example, if $(l,m,n)=(0,0,1)$, and if the spherical harmonics are defined with respect to this same axis, then the solution (\ref{404}) takes the form
\begin{equation}
\psi(\theta,\lambda,t)=-\Omega \sin \theta
+ \sum_{m=1}^n A_m P_n^m(\sin \theta) \cos \left( m(\lambda-\omega t) + \gamma_m \right)
\label{405}
\end{equation}
in spherical coordinates.
However, the axis $(l,m,n)$ is arbitrary, and, more importantly, the spherical harmonics can be defined with respect to any axis whatsoever;  it only matters that they are all of degree $n$.  In fact, the solution (\ref{404}), properly interpreted, holds in any system of coordinates. Viewed from a reference frame rotating with the solid-body flow---that is, viewed in rotating coordinates---the solution (\ref{404}) corresponds to a retrograde, `pseudo-westward', propagation of the $Y_n$-pattern at an angular velocity equal to the last term in (\ref{403}), which differs from the frequency of Rossby-Haurwitz waves in its lack of a factor $m$ in the numerator.  This $m$-factor is missing because we here define $\omega$ as angular frequency rather than wave frequency.  That is, in (\ref{405}) we write $m(\lambda-\omega t)$ rather than $m\lambda-\omega t$.

\cite{Kochin64} attribute the solution (\ref{404}) to Hans Ertel in 1945, but they do not give a reference.  See also \cite{Rochas84}. It was re-discovered by Herring's colleague \cite{Thompson82} for the special case in which $(l,m,n)$ coincides with the $z$-axis, as in (\ref{405}), and $Y_n$ consists of a \emph{single} spherical harmonic whose axis of symmetry is \emph{inclined} with respect to the $z$-axis by a prescibed angle.  Thompson proposed this special exact solution as a test of numerical codes based on spherical coordinates.  \cite{Verkley84} matched Thompson's solution in one region of the sphere with an analogous solution, in another region, in which $Y_n$ is replaced by the solutions of $\nabla_{LB}^2 Y_n =d Y_n$ with $d$ a \emph{positive} constant.  The matching conditions are that the two solutions precess about the $z$-axis at the same angular speed and have continuous derivatives at the boundary between regions.  The family of solutions discovered by Ertel, Thompson, and Verkley are noteworthy in that they seem to be the only analytic solutions of the vorticity equation on the sphere (besides `zonal' flows in which $\psi=F(\theta)$ with $F$ arbitrary) that do not include point vortices.  See also \cite{Neven92} and references therein.

We prove (\ref{404}) using stereographic coordinates.  Because rotations are involved, it is best to use the complex notation $\zeta=\xi+ i \eta$ and to regard $\psi(\xi,\eta,t)$, $q(\xi,\eta,t)$ as $\psi(\zeta,\bar{\zeta},t)$, $q(\zeta,\bar{\zeta},t)$, where $\bar{\zeta}=\xi- i \eta$ denotes the complex conjugate.
In this notation, 
\begin{equation} 
\nabla_{LB}^2 =(1+\zeta \bar{\zeta})^2  \frac{\partial^2 }{\partial \zeta \partial \bar{\zeta}}
\label{406}
\end{equation}
and the dynamics (\ref{28}) takes the form
\begin{equation}
h^2 q_t +2i \frac{\partial(\psi,q)}{\partial(\zeta,\bar{\zeta})}=0, \;\;\;\;
q
= \nabla_{LB}^2 \psi, \;\;\;\;
h=\frac{2}{1+\zeta \bar{\zeta}}
\label{407}
\end{equation}
given by \cite{Crowdy2004}.
Suppose, without loss of generality, that the axis of the solid body flow is $(0,0,1)$.  Then
\begin{equation}
\psi_{(0,0,1)}^\Omega= -\Omega \sin \theta = -\Omega z 
= -\Omega \frac{(\xi^2+\eta^2-1)}{(\xi^2+\eta^2+1)}
= -\Omega \frac{(\zeta \bar{\zeta}-1)}{(\zeta \bar{\zeta}+1)}
=\Omega h -\Omega.
\label{408}
\end{equation}
The last term in (\ref{408}) is an irrelevant constant.
Let $Y_n(\zeta,\bar{\zeta})$ be a solution of (\ref{401}).
To demonstrate Thompson's solution, we must show that
\begin{equation}
\psi=Y_n(s,\bar{s})+\Omega h
\label{409}
\end{equation} 
satisfies (\ref{407}), where
\begin{equation}
s \equiv \zeta e^{-i\omega t}, \;\;\;\; \bar{s} \equiv \bar{\zeta} e^{+i\omega t}
\label{410}
\end{equation}
and $\omega$ is given by (\ref{403}).  In the new coordinates $s,\bar{s}$,
\begin{equation} 
\nabla_{LB}^2 =(1+s\bar{s})^2  \frac{\partial^2 }{\partial s \partial \bar{s}}, \;\;\;\;
h=\frac{2}{1+s \bar{s}}.
\label{411}
\end{equation}
The vorticity corresponding to (\ref{409}) is
\begin{equation}
q= \nabla_{LB}^2 \left( Y_n(s,\bar{s})+\Omega h \right)
=-n(n+1) Y_n(s,\bar{s})-2\Omega h +2\Omega.
\label{412}
\end{equation}
The last term in (\ref{412}) is needed to satisfy the Gauss constraint, but it does not contribute to the vorticity equation, in which only derivatives of $q$ appear.  By the chain rule,
\begin{align}
h^2 q_t=-n(n+1) h^2 \frac{\partial}{\partial t} Y_n(s,\bar{s})
&=-n(n+1) i \omega h^2 \left( -s \frac{\partial Y_n}{\partial s}+\bar{s} \frac{\partial Y_n}{\partial \bar{s}} \right)
\notag \\
&=-2n(n+1)i\omega \frac{\partial(Y_n, h)}{\partial (s,\bar{s})}
\label{413}
\end{align}
where we have used the identities
\begin{equation}
\frac{\partial h}{\partial s}=-\frac{1}{2}h^2 \bar{s},  \;\;\;\; \frac{\partial h}{\partial \bar{s}}=-\frac{1}{2}h^2 s.
\label{414}
\end{equation}
For the second term in the vorticity equation we have
\begin{equation}
2i \frac{\partial(\psi,q)}{\partial(\zeta,\bar{\zeta})}=
2i \frac{\partial(Y_n+\Omega h, -n(n+1)Y_n-2\Omega h)}{\partial(s,\bar{s})}=
2\Omega i \left(n(n+1)-2\right) \frac{\partial(Y_n,h)}{\partial(s,\bar{s})}.
\label{415}
\end{equation}
The sum of (\ref{413}) and (\ref{415}) vanishes for $\omega$ given by (\ref{403}).  Thus (\ref{409}) satisfies our dynamics for any positive integer $n$. This proof is simpler than the proofs given by \cite{Thompson82} or \cite{Verkley84} using spherical coordinates.

\section{Numerical model}

We solve (\ref{27}-\ref{28}) by the method of \cite{Salmon89}, a generalization of \cite{Arakawa66}'s method that conserves discrete analogues of the energy (\ref{38}), total vorticity (\ref{39}), and enstrophy (\ref{40}) when $\nu=0$.  This method has been further generalized to the shallow water equations; see \cite{Salmon2009} and references therein.  

We begin by noting that the vorticity equation in (\ref{28}) is equivalent to the statement that
\begin{equation}
\iint d\xi d\eta \; \alpha h^2 q_t
= \iint d\xi d\eta \; \alpha \frac{\partial(\psi,q)}{\partial(\xi,\eta)}
\label{301}
\end{equation}
for any function $\alpha(\xi,\eta)$.  The integration is over the entire sphere.  By parts integrations, (\ref{301}) takes the more useful form
\begin{equation}
\iint d\xi d\eta \; \alpha h^2 q_t
= \frac{1}{3} \iint d\xi d\eta \; \left[ 
\alpha \frac{\partial(\psi,q)}{\partial(\xi,\eta)} + \psi \frac{\partial(q,\alpha)}{\partial(\xi,\eta)} 
+ q \frac{\partial(\alpha,\psi)}{\partial(\xi,\eta)} \right].
\label{302}
\end{equation}
Our strategy is to discretize all the variables in (\ref{302}) in such a way that the conservation of energy, total vorticity, and enstrophy correspond to purely algebraic cancellations between the terms.   The conservation of enstrophy  (\ref{40}) corresponds to the choice $\alpha=q$ in (\ref{302}). This requires that the discrete form of
\begin{equation}
\iint d\xi d\eta \; \left[ 
q \frac{\partial(\psi,q)}{\partial(\xi,\eta)} + \psi \frac{\partial(q,q)}{\partial(\xi,\eta)} 
+ q \frac{\partial(q,\psi)}{\partial(\xi,\eta)} \right]
\label{303}
\end{equation}
vanish, and this occurs automatically if the discrete Jacobian is antisymmetric.  Conservation of total vorticity corresponds to $\alpha=1$, and requires that the discrete form of
\begin{equation}
\iint d\xi d\eta \; 
\frac{\partial(\psi,q)}{\partial(\xi,\eta)}
\label{304}
\end{equation}
vanish.  Again, this occurs automatically.   Energy conservation corresponds to $\alpha=-\psi$, and we discuss it further below.

\begin{figure}[H]
\includegraphics[width=5. cm]{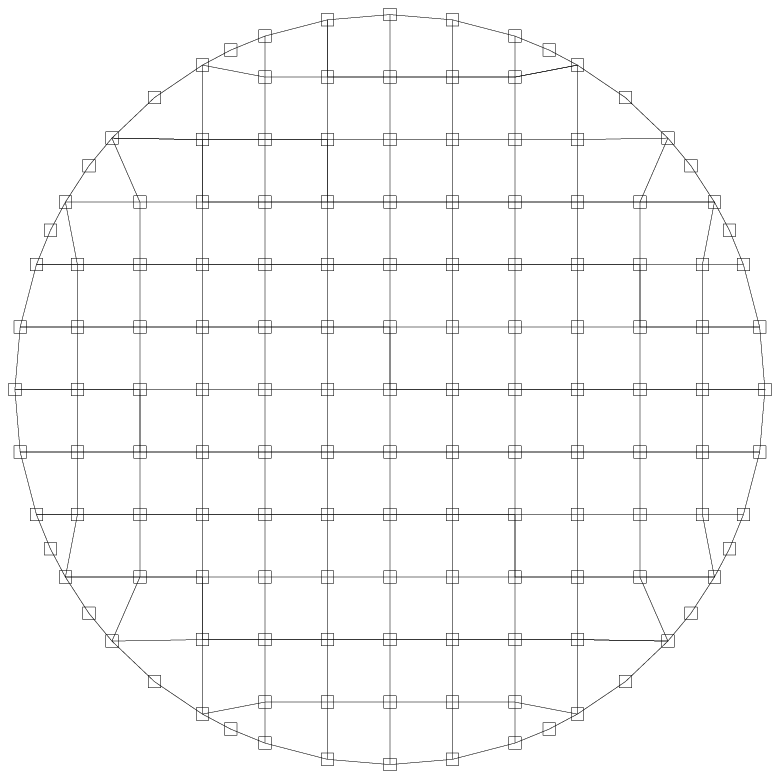}
\caption{Quadrilateral elements cover the interior of the unit circle in the stereographic plane.  The circle itself corresponds to the equator.  Most of the interior elements are perfect squares.  Elements with 1, 2, or 3 equatorial nodes are deformed quadrilaterals that conform to the circle.\label{fig2}}
\end{figure}

We discretize (\ref{302}) by covering the interior of both stereographic circles with quadrilateral finite elements.  
The placement of the elements is the same in both circles.
Most of the elements are perfect squares, as illustrated in Figure \ref{fig2} for a very low resolution case, but near the unit circle itself the quadrilaterals are deformed, with one or more nodes lying on the circle. 

The nodal values of $\psi$ and $q$ are the dependent variables of the model. 
Each interior node corresponds to a point on one of the hemispheres, and is shared by 4 elements.
Each equatorial node is shared by up to 6 elements, counting elements on both sides of the equator.  At every time, the state of the system is determined by the values of $\psi$ or $q$ at $N=2N_i+N_{eq}$ nodes, where $N_i$ is the number of interior nodes within each of the two unit circles, and $N_{eq}$ is the number of (shared) equatorial nodes.
Appendix B offers further details.  

In claiming conservation properties, we regard time derivatives as exact.  That is, we ignore errors that result from the finite time step.  Experience shows these errors to be small in comparison to those that result from space discretization.  This issue arises when we consider the finite-element discretization of the left-hand side of (\ref{302}), namely
\begin{equation}
\sum_i A_i \; \alpha_i h_i^2 \frac{dq_i}{dt}
\label{305}
\end{equation}
where the summation is over all $N$ nodes, $A_i$ is the area of the $\xi \eta$-plane associated with node $i$ (in a sense to be made precise), and $\alpha_i$, $h_i$, $q_i$ are the nodal values of $\alpha$, $q$ and $h$.  Note that the metric factor
\begin{equation}
h_i=\frac{2}{1+\xi_i^2+\eta_i^2}
\label{306}
\end{equation}
is determined by the arrangement of the quadrilateral elements, and is independent of time.
Replacing the left-hand side of (\ref{302}) with (\ref{305}), and the right-hand side of (\ref{302}) with the finite-element representation described briefly above and more thoroughly in Appendix B, we obtain the model dynamics by requiring that the discrete form of (\ref{302}) vanish for arbitrary nodal values $\alpha_i$.  Setting $\alpha_i=q_i$ and treating the time derivative as exact, we conclude that the discrete enstrophy
\begin{equation}
Z \equiv \frac{1}{2} \sum_i A_i \; h_i^2 q_i^2
\label{307}
\end{equation}
is conserved.

Energy conservation corresponds to the choice $\alpha_i=-\psi_i$.  Again the right-hand side of (\ref{302}) vanishes if the discrete Jacobian is antisymmetric, and hence
\begin{equation}
-\sum_i A_i \; \psi_i h_i^2 \frac{dq_i}{dt}=0.
\label{308}
\end{equation}
To make (\ref{308}) the time derivative of a discrete energy, we must consider the Poisson equation in (\ref{28}) relating $\psi$ and $q$.  It is equivalent to the statement that
\begin{equation}
-\iint d\xi d\eta \; \beta h^2 q
= \iint d\xi d\eta \; \nabla \beta \cdot \nabla \psi
\label{309}
\end{equation}
for any function $\beta(\xi,\eta)$.  The arbitrary function $\beta(\xi,\eta)$ is analogous to $\alpha(\xi,\eta)$.  
Let
\begin{equation}
-\sum_i A_i \; \beta_i h_i^2 q_i=\sum_{ij} w_{ij} \beta_i \psi_j
\label{310}
\end{equation}
be the discrete approximation to (\ref{309}), where $w_{ij}$ is the set of weights that arise by discretizing the right side of (\ref{309}).  
We require (\ref{310}) to vanish  for any set of $\beta_i$.
The $\beta_i$ are completely arbitrary, and can take different values at different times.  By first regarding the $\beta$ as time independent, we obtain the general result,
\begin{equation}
-\sum_i A_i \; \beta_i h_i^2 \frac{dq_i}{dt}=\sum_{ij} w_{ij} \beta_i \frac{d\psi_j}{dt}.
\label{311}
\end{equation}
Then, setting $\beta_i=\psi_i(t)$, treating time derivatives as exact, and assuming that the weights $w_{ij}$ are symmetric ($w_{ij}=w_{ji}$), we obtain
\begin{equation}
-\sum_i A_i \; \psi_i h_i^2 \frac{dq_i}{dt}=\sum_{ij} w_{ij} \psi_i \frac{d \psi_j }{dt}
=\frac{1}{2} \frac{d}{dt}\sum_{ij} w_{ij} \psi_i \psi_j .
\label{312}
\end{equation}
Since by (\ref{308}) this must vanish, we conclude that the discrete energy
\begin{equation}
E \equiv \frac{1}{2} \sum_{ij} w_{ij} \psi_i \psi_j 
\label{313}
\end{equation}
is conserved.

In overall summary, enstrophy is conserved if (\ref{302}) is discretized in such a way that its Jacobians are antisymmetric, and energy is conserved if (\ref{309}) is discretized in such a way that the dot product is symmetric. \cite{Salmon2009} shows how these two steps may be interpreted more generally in terms of the Poisson bracket and the Hamiltonian of the system.  The model does not conserve other Casimirs of the form (\ref{40a}).  Appendix B gives further details of the discretizations.

\section{Numerical solutions}

For flow with rms velocity $U$ on a sphere of radius $a$ with angular momentum corresponding to the solid body rotation rate $\Omega$, the Rossby number $Ro=U/\Omega a$ and the Rhines scale $Rh=Ro^{1/2} a$.  For values typical of the Earth's atmosphere---$U$=10 m/sec, $a$=6400 km, and $\Omega=2\pi$/day--- we have $Ro=0.02$ and $Rh=0.15 a$.  For Earth's ocean and the giant planets the ratio of the Rhines $Rh$ scale to the planetary radius $a$ is even smaller.  This regime, which features zonal jets with widths on the order of $Rh$, has attracted the greatest interest from modelers, but the non-rotating case $\Omega=0$ is also interesting, because it reveals the effects of curvature by itself.

In our experiments $U=a=1$, where $U$ is the rms velocity at the intial time in the rotating frame, that is, the rms velocity excluding solid body rotation.  The Rossby number $Ro=0.02$ corresponding to the Earth's atmosphere then corresponds to $\Omega=50$. We present solutions of the two cases $\Omega=0$ and $\Omega=50$.  In the solutions with $\Omega=50$, the displayed stream function and vorticity do not include the $n=1$ modes associated with the angular momentum and therefore depict the fields in the rotating frame. In all of our experiments, the viscosity, if nonzero, has the value $\nu=q_{rms}\Delta^2$, where $q_{rms}$ is the rms vorticity and $\Delta^2$ is the area of the square gridboxes on the stereographic plane.  The number $N=241,000$ of nodes corresponds to the cutoff $n_c=490$ in spherical harmonic degree $n$.

\begin{figure}[H]
\includegraphics[width=11. cm]{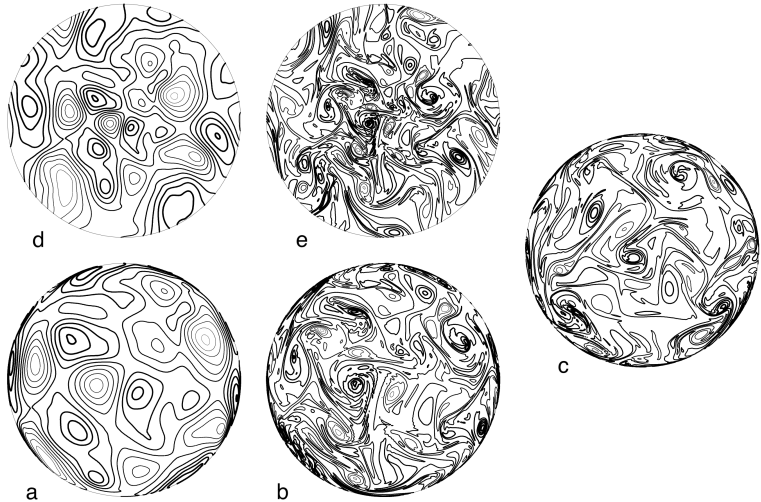}
\caption{Stream function, (a) and (d), and vorticity, (b), (c) and (e), at time $t=2.0$ in a non-rotating, viscous experiment in which the initial energy is confined to spherical harmonic degrees $n=16,17,18$. (a) and (b) are polar views, as seen by an observer at $z=\infty$, above the north pole. (d) and (e) are northern-hemisphere stereographic views that depict the same field as the polar views directly below.  Compared to the polar views (a) and (b), the stereographic views (d) and (e) are relatively free of distortion.  Panel (c) is an equatorial view, as seen by an observer at $x=\infty$.  In all views, darker contour lines correspond to larger values.}
\label{fig3}
\end{figure}

\begin{figure}[H]
\includegraphics[width=11. cm]{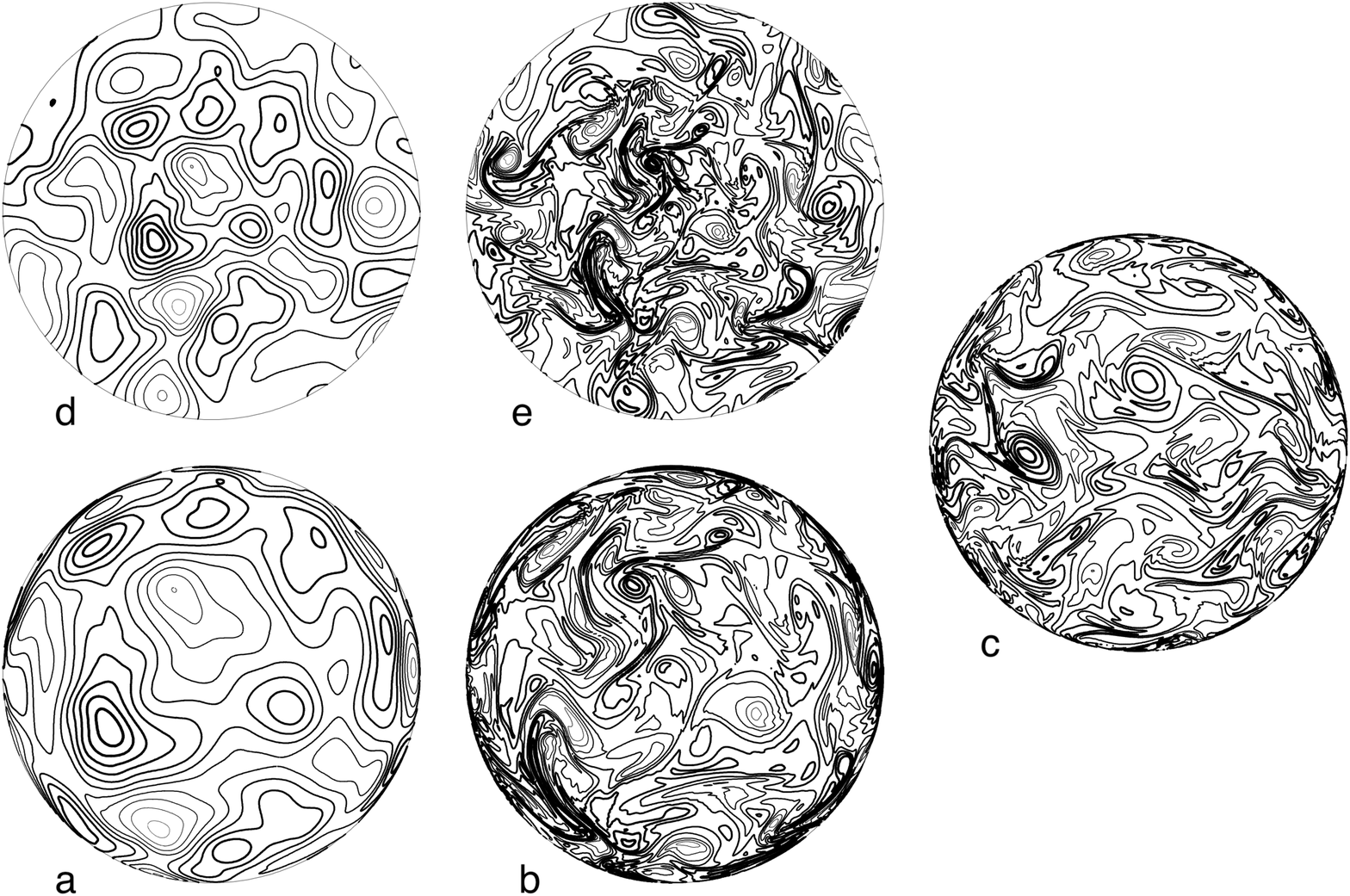}
\caption{The same as Figure \ref{fig3}, but for the case $\Omega=50$ of Earth-like rotation.  This solution differs from that of Figure \ref{fig3} in that isolated coherent vortices are less prominent, and that elongated, jet-like structures are beginning to appear, especially near the equator in (c), where the $\beta$-effect is greatest.}
\label{fig4}
\end{figure}

Figure \ref{fig3} shows the streamfunction and vorticity in 3 views of the sphere at time $t=2.0$ in an experiment with $\Omega=0$ (the non-rotating case).  The initial energy is confined to spherical harmonics of degree $n=16,17,18$ with the amplitudes randomly assigned.  Figure \ref{fig4} shows the same fields in an experiment beginning from the same initial conditions but with $\Omega=50$.

In Figures \ref{fig3} and \ref{fig4}, (a) is a polar view of the stream function in the northern hemisphere, as seen by an observer at $z=\infty$ above the north pole, and (d) is the same field within the stereographic cirle $\hat{\xi}^2+\hat{\eta}^2 <1$ corresponding to the northern hemisphere.  Although both views cover the entire northern hemisphere, the polar view (a) severely distorts features near the equator: The ratio between true distance and apparent distance in (a) varies between unity at the pole and infinity at the equator. In contrast, the stereographic view (d) introduces a scale distortion that varies only between $1$ and $2$.  It offers the superior view of the entire northern hemisphere field.

Figure \ref{fig4} shows the same 5 views as in Figure \ref{fig3}, but for the case $\Omega=50$ of nonvanishing angular momentum. Figure \ref{fig4} differs from Figure \ref{fig3} in that its isolated coherent vortices are less prominent than in Figure \ref{fig3}, and elongated, jet-like structures are beginning to appear, especially near the equator (Figure \ref{fig4}c), where the $\beta$-effect is greatest.

Figures \ref{fig3} and \ref{fig4} illustrate the advantage of stereographic projection.  Whereas at least six conventional views of the sphere are needed to give a fair representation of a single field, two stereographic projections offer a complete and relatively undistorted depiction. 

\begin{figure}[H]
\includegraphics[width=15. cm]{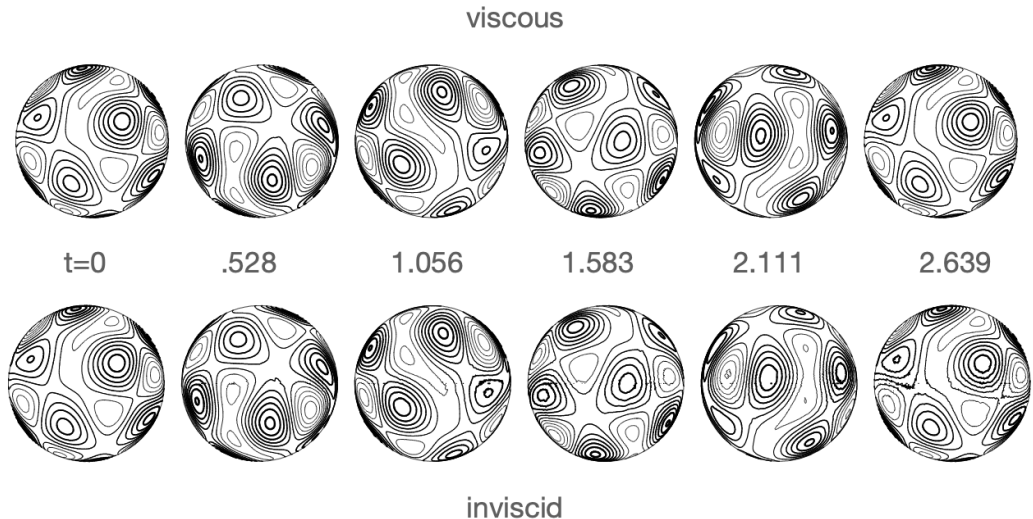}
\caption{Vorticity as seen from $x=\infty$ at 6 times in two solutions in which the angular momentum vector points in the $y$-direction.  $z$ increases upward in the figure.  The initial state at $t=0$ consists only of harmonics of degree $n$.  The pattern rotates upward in the figure, around the sphere, to return to its original location at the predicted time $t=2.639$.  The inviscid solution shows small, grid-scale oscillations in the vorticity that do not appear in the viscous solution.
\label{fig5}}
\end{figure}

Next we consider two experiments in which the initial state is sharply confined to spherical harmonics of degree $n=6$. Angular momentum corresponding to a solid body rotation rate $\Omega=50$ is present, but now it points in the direction of the $y$-axis in the Cartesian embedding space. This is achieved by replacing the Coriolis parameter $f$ in (\ref{29a}) by
\begin{equation}
2\Omega y =2\Omega \frac{2\eta}{\xi^2+\eta^2+1}
=2\Omega\frac{2\hat{\eta}}{\hat{\xi}^2+\hat{\eta}^2+1}.
\label{501}
\end{equation}
According to Section 3, the $n=6$ initial state should rotate rigidly, in a retrograde manner, about the $y$-axis at the angular rate $\omega\equiv2\Omega/(6(6+1))$. Figure \ref{fig5} shows this motion as seen by an observer at $x=\infty$.  From its perspective, the pattern rotates upward in the figure, around the sphere, completing a full cycle of revolution in the recurrence time $2\pi/\omega=2.639$ shown on the far right of the figure.  The solutions depicted in Figure \ref{fig5} are both a check on the theory and a strong test of the numerical code, because the rotation carries the pattern from the interior of one stereographic circle to the other and then back again, twice crossing the irregularly shaped elements near the bounding circles (\ref{7}). The inviscid experiment (bottom row of Figure \ref{fig5}) exhibits small, grid-scale oscillations in the vorticity field, which do not appear in the viscous experiment (top row).

Finally, we explore the suggestion in Section 1 that solutions lock into a static state as Bose-Einstein condensation occurs.  Theory predicts Bose-Einstein condensation only as $n_c\to \infty$ and $t\to \infty$---two limits that are impossible to achieve numerically. The interesting question is whether quasi-static states appear with finite resolution and in finite time.  According to equilibium statistical mechanics \citep{Frederiksen80}, the equilibrium energy in spherical harmonic degree $n$ is
\begin{equation}
E_n=\frac{2n+1}{\alpha +\beta n(n+1)}, \;\;\;\; n\ge 2
\label{502}
\end{equation}
where the constants $\alpha$ and $\beta$ are determined by the energy and enstrophy. The equilibrium spectrum (\ref{502}) is the analogue of \cite{Kraichnan1967}'s  spectrum
\begin{equation}
E(k)=\frac{k}{\alpha +\beta k^2}
\label{503}
\end{equation}
for flow on the plane.  The numerator in (\ref{502}) is the number of modes of degree $n$. Bose-Einstein condensation corresponds to the limit $\alpha \to -2(2+1)\beta$ as $n_c\to \infty$.

Several authors investigate freely decaying turbulence on the sphere and its resemblance to the predictions of equilibrium statistical mechanics, including the more recondite Miller-Sommeria-Robert-Montgomery (MRSM) theory. See
\cite{Qi2014},
\cite{Dritschel2015},
\cite{Modin2020}, and
\cite{Jagad2021}.
For a summary of the MRSM theory, see \cite{Eyink2006}.  The requirement that $t\to \infty$ has proved formidable. MRSM theory claims to predict the structure of coherent vortices by the principle of maximum entropy. \cite{Salmon2018} offers a different picture, in which coherent vortices are \emph{low} entropy sites that exist in order to maximize the \emph{rate} of entropy production in the area outside the vortices.

\begin{figure}[H]
\includegraphics[width=10. cm]{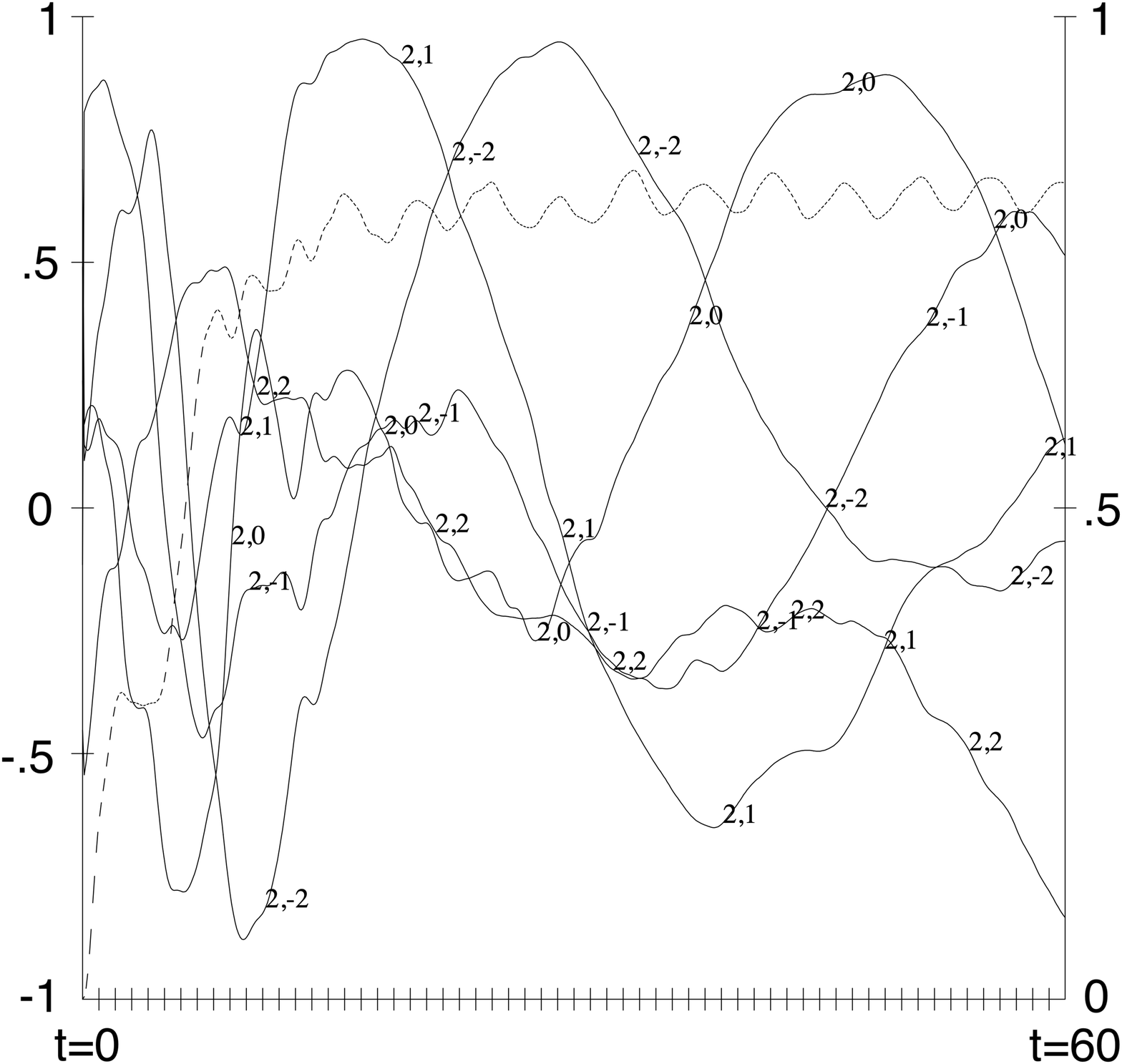}
\caption{ Evolution of a non-rotating, inviscid flow in which the initial energy is confined to spherical harmonic degrees $n=4,5,6$. The dashed curve (right scale) is the fraction of energy in $n=2$, which vanishes initially, but, according to theory, will eventually reach 0.996.  The other curves, labeled $2,m$, represent the normalized coefficient of $Y_2^m$ in the spherical harmonic representation of the flow, which vary between $\pm 1$ (left scale).}
\label{fig6}
\end{figure}

Here we consider the case $\Omega=\nu=0$ of nonrotating, inviscid flow with a modest resolution corresponding to $n_c=240$.  The initial condition corresponds to energy distributed equally among the three degrees $n=4,5,6$, that is, $E_4=E_5=E_6$ with all other $E_n=0$.  The amplitudes of the spherical harmonics within each band are randomly assigned, and the initial state is normalized such that $u_{rms}=1.$ A calculation based upon (\ref{502}) predicts that 99.6\% of the energy will end up in $n=2$, the lowest available mode.  Again, $E_1=0$ at all time by the conservation of angular momentum. Figure \ref{fig6} summarizes the evolution of the flow from $t=0$ to $t=60$.  The time $t=60$ is the time required for a fluid particle to circumnavigate a great circle $60/2\pi=9.5$ times at the rms velocity of the flow.  The dashed curve in Figure \ref{fig6} is the fraction of energy in $n=2$.  Beginning from zero, it only gradually approaches its predicted value of 0.996.

The solid curves in Figure \ref{fig6} measure the approach to a static configuration.  Let $A_{2,m}$ be the amplitude of $Y_2^m$ in the spherical harmonic representation of $\psi$, and define
\begin{equation}
\mathbf{p}=\frac{(A_{2,0},\;A_{2,1},\;A_{2,-1},\;A_{2,2},\;A_{2,-2})}
{\left[ (A_{2,0})^2 +(A_{2,1})^2 +(A_{2,-1})^2 +(A_{2,2})^2 +(A_{2,-2})^2 \right]^{1/2}}.
\label{504}
\end{equation}
Thus $\mathbf{p}$ is a unit vector in the 5-dimensional space spanned by the amplitudes of the 5 spherical harmonics comprising $n=2$.  If the flow were indeed to become static, then the five components of $\mathbf{p}$ would become time-independent at values that are accidents of the initial conditions.  Our choice of conventionally defined spherical harmonics as amplitudes to be tested is both arbitrary and irrelevant; any other choice of mutually orthogonal functions that span $\psi_2$ would serve as well.

Figure \ref{fig6} shows that the approach to the putative equilibrium state is extremely slow.  Although $E_2$ generally increases, its increase is not monotonic, but instead exhibits persistent, relatively rapid oscillations.  The components of $\mathbf{p}$, labeled $n,m$ in the figure oscillate over their full range of $\pm 1$, but the periods of their oscillations appear to lengthen as time increases.  It is conceivable that this flow eventually reaches a static state, but this evidently requires a very long time.  On the other hand, since the time unit is the time required for a fluid particle to move a distance equal to the radius of the sphere, the flow at $t=60$ is already quasi-static in the sense that it requires at least several such time units for the components of $\mathbf{p}$ to change appreciably.

\appendix
\section[\appendixname~\thesection]{Viscosity in stereographic coordinates.}

The angular-momentum-conserving viscosity suggested by \cite{Gilbert2014} is
\begin{equation}
\frac{\partial \dot{\xi}^i}{\partial t}+ \cdots = \nu \left( g^{kl} \nabla_k \nabla_l \dot{\xi}^i -K\dot{\xi}^i \right)
\tag{A1}
\end{equation}
where $(\xi^1,\xi^2)$ are general coordinates covering the sphere, $\dot{\xi}^i=d\xi^i/dt$ following a fluid particle, $g^{kl}$ is the inverse of the metric tensor $g_{kl}$, $\nabla_k$ is the covariant derivative, and $K$ is the Gaussian curvature.  In this appendix, we use raised indices to denote contravariant objects and lowered indices to denote covariant objects, as is customary in tensor analysis.  Repeated indices are summed.  For the unit sphere $K=1$.  The ellipses denote the non-viscous terms that are not at issue here.
In the stereographic coordinates $(\xi,\eta)$, this viscosity law takes the form
\begin{equation}
\frac{\partial u^i}{\partial t}+ \cdots = 
\nabla_{\xi} \nabla_{\xi} (\gamma^2 u^i)  
+\nabla_{\eta} \nabla_{\eta} (\gamma^2 u^i)     -u^i
\tag{A2}
\end{equation}
where 
$(u^1,u^2)=(u,v)=h^2(\dot{\xi},\dot{\eta})$ as previously defined, and $\gamma \equiv 1/h$.   To ease the notation, we have set $\nu=1$.

For a general contravariant vector $w^i$
\begin{equation}
\nabla_k w^i=\frac{\partial w^i}{\partial x^k}+\Gamma^i_{jk}w^j
\tag{A3}
\end{equation}
where
\begin{equation}
\Gamma^i_{jk}=
\frac{1}{2} g^{il}\left( \frac{\partial g_{lj}}{\partial x^k}+\frac{\partial g_{lk}}{\partial x^j}-\frac{\partial g_{kj}}{\partial x^l} \right).
\tag{A4}
\end{equation}
For mixed tensors $T^i_{k}$, such as $\nabla_k w^i$
\begin{equation} 
\nabla_l T^i_{k}= \frac{\partial T^i_{k}}{\partial x^l} +\Gamma^i_{k\mu}T^{\mu}_l-\Gamma^{\mu}_{kl}T^i_{\mu}.
\tag{A5}
\end{equation}
In stereographic coordinates, the non-vanishing metric coefficients are
\begin{equation}
g_{\xi \xi}=g_{\eta \eta}= g, \;\;\; g^{\xi \xi}=g^{\eta \eta}= \frac{1}{g}
\tag{A6}
\end{equation}
where
\begin{equation}
g\equiv h^2 = \frac{4}{\left( 1+\xi^2+\eta^2\right)^2}
\tag{A7}
\end{equation}
and the corresponding Christoffel symbols are
\begin{equation}
\Gamma^\xi_{\xi \xi}=-\Gamma^\xi_{\eta \eta}=\Gamma^\eta_{\xi \eta}
=-h\xi
\tag{A8}
\end{equation}
and
\begin{equation}
\Gamma^\eta_{\eta \eta}=-\Gamma^\eta_{\xi \xi}=\Gamma^\xi_{\eta \xi}
=-h\eta.
\tag{A9}
\end{equation}
The needed second covariant derivatives are of the form
\begin{equation} 
\nabla_\xi \nabla_\xi w^i=
\frac{\partial}{\partial \xi} \nabla_\xi w^i  
+\Gamma^i_{\xi \mu}\nabla_\xi w^{\mu}
-\Gamma^{\mu}_{\xi \xi}\nabla_\mu w^i
\tag{A10}
\end{equation}
and 
\begin{equation} 
\nabla_\eta \nabla_\eta w^i=
\frac{\partial}{\partial \eta} \nabla_\eta w^i  
+\Gamma^i_{\eta \mu}\nabla_\eta w^{\mu}
-\Gamma^{\mu}_{\eta \eta}\nabla_\mu w^i
\tag{A11}
\end{equation}
where $(w^{\xi},w^{\eta})\equiv \gamma^2 (u,v)= (\dot{\xi},\dot{\eta})$.
Using (A6-9), we obtain
\begin{align}
\nabla_\xi \nabla_\xi w^\xi&=\frac{\partial}{\partial \xi}\nabla_\xi w^\xi
-h\eta \left( \nabla_\xi w^\eta + \nabla_\eta w^\xi \right)
\tag{A12} \\
\nabla_\xi \nabla_\xi w^\eta&=\frac{\partial}{\partial \xi}\nabla_\xi w^\eta
+h\eta \left( \nabla_\xi w^\xi - \nabla_\eta w^\eta  \right)
\tag{A13} \\
\nabla_\eta \nabla_\eta w^\xi&=\frac{\partial}{\partial \eta}\nabla_\eta w^\xi
+h\xi \left( \nabla_\eta w^\eta - \nabla_\xi w^\xi  \right)
\tag{A14} \\
\nabla_\eta \nabla_\eta w^\eta&=\frac{\partial}{\partial \eta}\nabla_\eta w^\eta
-h\xi \left( \nabla_\eta w^\xi + \nabla_\xi w^\eta \right)
\tag{A15}
\end{align}
for the second covariant derivatives.  
For the first covariant derivatives we obtain
\begin{align}
\nabla_\xi w^\xi&=\frac{\partial w^\xi}{\partial \xi}-h\xi w^\xi-h\eta w^\eta
\tag{A16} \\
\nabla_\xi w^\eta&=
\frac{\partial w^\eta}{\partial \xi} +h\eta w^\xi-h\xi w^\eta
\tag{A17} \\
\nabla_\eta w^\xi&=
\frac{\partial w^\xi}{\partial \eta} -h\eta w^\xi+h\xi w^\eta
\tag{A18}\\
\nabla_\eta w^\eta&=
\frac{\partial w^\eta}{\partial \eta} -h\xi w^\xi-h\eta w^\eta.
\tag{A19}
\end{align}
By (A17-18)
\begin{equation}
\nabla_\xi w^\eta+\nabla_\eta w^\xi =
\frac{\partial w^\eta}{\partial \xi} + \frac{\partial w^\xi}{\partial \eta}.
\tag{A20}
\end{equation}
The incompressibility condition
\begin{equation}
\nabla_\xi w^\xi +\nabla_\eta w^\eta = 0
\tag{A21}
\end{equation}
implies that
\begin{equation}
h\xi w^\xi + h\eta w^\eta = \frac{1}{2} \left( \frac{\partial w^\xi}{\partial \xi} + \frac{\partial w^\eta}{\partial \eta} \right)
\tag{A22}
\end{equation}
so that (A16) and (A19) can be written as
\begin{align}
\nabla_\xi w^\xi&=\frac{1}{2} \left( \frac{\partial w^\xi}{\partial \xi} - \frac{\partial w^\eta}{\partial \eta}\right)
\tag{A23} \\
\nabla_\eta w^\eta&=\frac{1}{2} \left( \frac{\partial w^\eta}{\partial \eta} - \frac{\partial w^\xi}{\partial \xi} \right).
\tag{A24}
\end{align}
Thus
\begin{equation}
\nabla_\xi w^\xi - \nabla_\eta w^\eta= \left( \frac{\partial w^\xi}{\partial \xi} - \frac{\partial w^\eta}{\partial \eta}\right).
\tag{A25}
\end{equation}
Using these equations to simplify the needed second covariant derivatives (A12-15) we obtain
\begin{align}
\nabla_\xi \nabla_\xi w^\xi&=\frac{1}{2}\frac{\partial}{\partial \xi}\left( \frac{\partial w^\xi}{\partial \xi} - \frac{\partial w^\eta}{\partial \eta}\right)
-h\eta \left( \frac{\partial w^\eta}{\partial \xi} + \frac{\partial v^\xi}{\partial \eta}\right)
\tag{A26} \\
\nabla_\xi \nabla_\xi w^\eta&=\frac{\partial}{\partial \xi}\left( \frac{\partial w^\eta}{\partial \xi} +h\eta w^\xi-h\xi w^\eta \right)
+h\eta \left( \frac{\partial w^\xi}{\partial \xi} - \frac{\partial w^\eta}{\partial \eta}  \right)
\tag{A27} \\
\nabla_\eta \nabla_\eta w^\xi&=\frac{\partial}{\partial \eta}\left( \frac{\partial w^\xi}{\partial \eta} -h\eta w^\xi+h\xi w^\eta \right)
+h\xi \left( \frac{\partial w^\eta}{\partial \eta} - \frac{\partial w^\xi}{\partial \xi} \right)
\tag{A28} \\
\nabla_\eta \nabla_\eta w^\eta&=\frac{1}{2}\frac{\partial}{\partial \eta} \left( \frac{\partial w^\eta}{\partial \eta} - \frac{\partial w^\xi}{\partial \xi} \right)
-h\xi \left( \frac{\partial w^\eta}{\partial \xi} + \frac{\partial w^\xi}{\partial \eta}\right).
\tag{A29}
\end{align}
Finally, setting $(w^{\xi},w^{\eta})=\gamma^2(u,v)$, substituting (A26-29) into (A2), and simplifying the resulting expressions, we obtain the result (\ref{42}-\ref{43}).

\section[\appendixname~\thesection]{Finite-element discretization.}

\begin{figure}[H]
\includegraphics[width=5. cm]{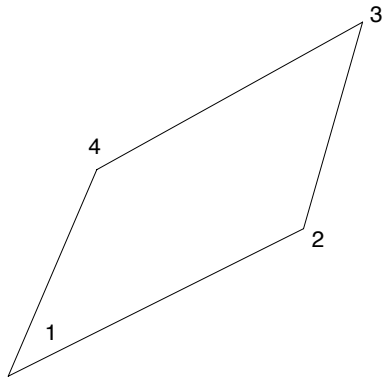}
\caption{A representative quadrilateral element with nodes labeled 1 to 4.\label{figB1}}
\end{figure}

Every finite element is a quadrilateral.  Most of the quadrilaterals are perfect squares, but those near the unit circles are deformed.  Figure \ref{figB1} shows a representative element with nodes numbered 1 to 4. 

Let $\{\gamma_i, \; i=1,4\}$ be the nodal values of $\gamma(\xi,\eta)$, which could be $\beta(\xi,\eta)$, $\psi(\xi,\eta)$, or either of the coordinates $\xi$ or $\eta$ itself.  Within each element
\begin{equation}
\gamma(\xi,\eta)=\sum_{i=1}^4 \gamma_i N_i(s,t)
\tag{B1}
\end{equation}
where $(s,t)$ are `labeling' coordinates that cover the element in the domain $-1<s,t<1$;
\begin{equation}
N_i(s,t)=\frac{1}{4}(1+ss_i)(1+tt_i), \;\;\;\; i=1 \ldots 4
\tag{B2}
\end{equation}
is the standard shape function for four-node quadrilateral finite elements; and
\begin{equation}
(s_1,t_1)=(-1,-1), \;\;\; (s_2,t_2)=(1,-1), \;\;\; (s_3,t_3)=(1,1), \;\;\; (s_4,t_4)=(-1,1)
\tag{B3}
\end{equation}
correspond, respectively, to the corners numbered 1, 2, 3, 4 in Figure \ref{figB1}. Thus $N_i=1$ at node $i$ and $N_i=0$ at the other three nodes.
(In this appendix, the symbol $t$ has nothing to do with time.)

The contribution of a single element to the right-hand side of (\ref{309}) is
\begin{equation}
\iint d\xi d\eta \; \nabla \beta \cdot \nabla \psi
=\int_{-1}^1 \int_{-1}^1 ds dt \; \frac{\partial(s,t)}{\partial(\xi,\eta)} \left[
\frac{\partial(\beta,\eta)}{\partial(s,t)}   \frac{\partial(\psi,\eta)}{\partial(s,t)}  +
\frac{\partial(\xi,\beta)}{\partial(s,t)}    \frac{\partial(\xi,\psi)}{\partial(s,t)}  \right].
\tag{B4}
\end{equation}
The first factor in the integrand is the ratio of quadrilateral area on the $st$-plane, namely 4, to the corresponding area $A_{1234}$ on the $\xi \eta$-plane, which is conveniently calculated by bisecting the deformed quadrilateral into two triangles.  Thus
\begin{equation}
\iint d\xi d\eta \; \nabla \beta \cdot \nabla \psi
=\frac{4}{A_{1234}}  \iint ds dt \;  \left[
\frac{\partial(\beta,\eta)}{\partial(s,t)}   \frac{\partial(\psi,\eta)}{\partial(s,t)}  +
\frac{\partial(\xi,\beta)}{\partial(s,t)}    \frac{\partial(\xi,\psi)}{\partial(s,t)}  \right].
\tag{B5}
\end{equation}
For the first Jacobian in (B5) we obtain
\begin{equation}
\frac{\partial(\beta,\eta)}{\partial(s,t)}=\sum_{i,j=1,4} \beta_i \eta_j F_{ij}(s,t)
\tag{B6}
\end{equation}
where
\begin{equation}
F_{ij}(s,t)=\frac{1}{16}\left[ s_it_j(1+tt_i)(1+ss_j)-s_jt_i(1+tt_j)(1+ss_i) \right].
\tag{B7}
\end{equation}
Substituting (B6), (B7) and the corresponding expressions for the other Jacobians back into (B5), we obtain
\begin{equation}
\iint d\xi d\eta \; \nabla \beta \cdot \nabla \psi
=\frac{4}{A_{1234}}  \sum_{i,j,n,m=1}^4 \Gamma_{ijnm} \left[
\beta_i \eta_j \psi_n \eta_m +\xi_i\beta_j\xi_n\psi_m  \right]
=  \sum_{i,j=1}^4 \beta_i C_{ij} \psi_j
\tag{B8}
\end{equation}
where
\begin{equation}
\Gamma_{ijnm}=\int_{-1}^1 \int_{-1}^1 dsdt \; F_{ij}(s,t) F_{nm}(s,t)
\tag{B9}
\end{equation}
and
\begin{equation}
C_{ij}= \frac{4}{A_{1234}} \sum_{pq} \left[
\eta_p\eta_q\Gamma_{ipjq}+\xi_p\xi_q\Gamma_{piqj} \right].
\tag{B10}
\end{equation}
The $C_{ij}$ may be calculated analytically computed and stored. 

The final sum in (B8) represents the contribution of a single element.  To discretize the integral (B4) over the entire sphere, we sum over all the elements to obtain (\ref{310}).    The weight $w_{ij}$ vanishes unless node $i$ and node $j$ share at least one element.  Thus, up to 6 elements contribute to each $w_{ij}$.  However, since $C_{ij}$ is symmetric for each contributing element, $w_{ij}$ is symmetric as well, and therefore the discrete energy (\ref{313}) is conserved.

The right-hand side of (\ref{302}) is
\begin{equation}
 \frac{1}{3} \iint d\xi d\eta \; \left[ 
\alpha \frac{\partial(\psi,q)}{\partial(\xi,\eta)} + cyc\{\alpha,\psi,q\} \right]
\tag{B11}
\end{equation}
where the integral is over the whole sphere.
We discretize it in a similar manner to (B8).
For the integral over the single element depicted in Figure \ref{figB1}, we approximate
\begin{equation}
\iint d\xi d\eta \; 
\alpha \frac{\partial(\psi,q)}{\partial(\xi,\eta)}
=\left( \frac{1}{4}\sum_{i=1}^4 \alpha_i \right) \iint d\xi d\eta \; \frac{\partial(\psi,q)}{\partial(\xi,\eta)}
\tag{B12}
\end{equation}
and evaluate
\begin{equation}
\iint d\xi d\eta \; \frac{\partial(\psi,q)}{\partial(\xi,\eta)}
= \oint q \frac{\partial \psi}{\partial s} ds
=\frac{1}{2} \sum_{j=1}^4 q_j (\psi_{j+1}-\psi_{j-1})
\tag{B13}
\end{equation}
where $s$ is distance on the $\xi \eta$-plane, measured counter-clockwise around the perimeter of the element.  In the sum over $j$, if $j=1$ then $j-1=4$, and if $j=4$ then $j+1=1$.  Thus the contribution of each element to (\ref{302}) is
\begin{equation}
 \frac{1}{24}  \sum_{i=1}^4 \alpha_i \sum_{j=1}^4  q_j (\psi_{j+1}-\psi_{j-1})  + cyc\{\alpha,\psi,q\}.
 \tag{B14}
\end{equation}
Summing over all the elements gives the discrete approximation to (\ref{302}).  

The left-hand sides of (\ref{301}) and (\ref{309}) take the discrete form
\begin{equation}
\sum_i A_i \; \alpha_i h_i^2  f_i
\tag{B15}
\end{equation}
where $f_i=dq_i/dt$ in the case of (\ref{301}) and $f_i=q_i$ in the case of (\ref{309}).  In (B15) the summation is over nodes, and $A_i$ is the area of the $\xi \eta$-plane that is `assigned' to, i.e. closest to, node $i$.  The precise rule of assignment is arbitrary, but it must obey the normalization requirement that the sum of the $A_i$'s equals the area within a unit circle, or that the sum of the $h_i^2 A_i$'s on both unit circles equals the area of the unit sphere.

By requiring the coefficient of each $\beta_i$ in (\ref{310}) to vanish, we obtain $N$ equations representing a logical discrete approximation  to $h^2 q =\psi_{\xi \xi}+\psi_{\eta \eta}$.  
By requiring the coefficient of each $\alpha_i$ in the above-described discrete analogue of (\ref{28}) to vanish, we obtain $N$ equations representing a logical discrete approximation  to (\ref{28}).



\bibliography{ref}
\bibliographystyle{apalike}
\end{document}